\pdfoutput=1
\documentclass[11pt, a4paper]{article}
\usepackage{graphicx} 
\usepackage{subcaption}
\usepackage[utf8]{inputenc}
\usepackage{vmargin}
\setmargins{2.5cm}{1.5cm}{16.5cm}{23.42cm}{10pt}{1cm}{0pt}{2cm}
\usepackage{mathtools, amssymb, amsfonts, amsthm, bbm}
\usepackage{microtype}
\newtheorem{theorem}{Theorem}

\newtheorem{lemma}[theorem]{Lemma}
\newtheorem{proposition}{Proposition}
\newtheorem{remark}{Remark}
\theoremstyle{definition}
\newtheorem{definition}{Definition}

\usepackage{notation}
\newcommand{\Qlambda}{\mathcal{Q}_{\lambda}^{d}}

\usepackage{enumitem}
\usepackage{cite}
\begin{document}
\begin{center}
    {\bfseries \LARGE On Universal Quantum Hypothesis Testing} \\
    \vspace{5mm}
    {\Large Cristian Ferrezuelo-Oteo, Gonzalo Vazquez-Vilar, and Tobias Koch} \\
    \vspace{2mm}
    {\small Signal Theory and Communications Department, Universidad Carlos III de Madrid, 28911 Leganés, Spain} \\
    {\small Gregorio Mara\~n\'on Health Research Institute, 28007 Madrid, Spain} \\
    {\small Emails: cferrezu@pa.uc3m.es, gonzalo.vazquez@uc3m.es, tkoch@ing.uc3m.es}
\end{center}
\begingroup
\renewcommand\thefootnote{}%
\footnote{The authors have received funding from the Spanish Ministerio de Ciencia, Innovaci\'on y Universidades under Grant PID2024-159557OB-C21 (MICIU/AEI/10.13039/501100011033 and ERDF/EU), and from the Comunidad de Madrid under Grants IDEA-CM (TEC-2024/COM-89) and PEJ-2024-AI/COM-32104, the latter co-funded by the European Union through the European Social Fund Plus (ESF+).}%
\addtocounter{footnote}{-1}%
\endgroup
\vspace{1cm}
\begin{abstract}
  This paper studies quantum one-sample testing, a universal quantum hypothesis testing problem in which a sequence of tests must distinguish a known null state from an arbitrary, unknown alternative state. When both the null state and the alternative state are known, the optimal type-II error exponent in Stein's regime is given by the Umegaki relative entropy between the states. Here, it is demonstrated that, when the alternative state is unknown, there is no sequence of universal tests for which this exponent can be achieved for all alternative states, unless the null state is maximally mixed. This implies that there is no sequence of tests that is universally optimal. The paper further proposes a sequence of universal tests that is based on tilted POVMs in the Schur--Weyl decomposition. These tests achieve a type-II error exponent in Stein's regime that is lower-bounded by the reverse sandwiched relative entropy between the null and alternative states.
\end{abstract}
\section{Introduction}\label{sec:intro}
Quantum hypothesis testing is a fundamental task in quantum information theory. In the binary setting, given $n$ copies of a quantum state, the goal is to discriminate between a null hypothesis and an alternative hypothesis. A test is a decision rule, based on measurements of the available copies, that decides which hypothesis is true. There are two types of errors: a type-I error consists of rejecting the null hypothesis when it is true, whereas a type-II error consists of accepting the null hypothesis when the alternative is true. We consider the asymptotic limit, where a large number of independent and identically distributed (i.i.d.) copies of the state are available. We further consider the asymmetric Stein regime, where the type-I error probability is bounded above by a fixed threshold and the type-II error probability is required to decay exponentially with $n$.

When a hypothesis comprises a single quantum state, it is called \emph{simple}. Binary asymmetric hypothesis testing between a simple null hypothesis and a simple alternative hypothesis is well understood. For every $n$, the quantum Neyman--Pearson test minimizes the type-II error subject to a bound on the type-I error \cite{QuantumNeymanPearson}. In the asymptotic limit, the quantum Stein lemma characterizes the Umegaki quantum relative entropy \cite{Umegaki_relative_entropy} as the optimal type-II error exponent under any fixed or vanishing type-I error threshold \cite{ProperFormulaRelativeEntropy, Strong_converse_and_Stein_s_lemma_in_quantum_hypothesis_testing}. This result is the quantum counterpart of the classical Chernoff--Stein lemma \cite{Stein_lemma_unpublished, Chernoff_article}, where the optimal error exponent is the relative entropy between the probability distributions of the null and alternative hypotheses; see also \cite[Section 11.8]{ElementsOfInformationTheoryBook}.

In composite quantum hypothesis testing, one or both hypotheses are sets of states. The test determines whether the unknown state belongs to a null set $\Nc$ or an alternative set $\Ac$. Under the minimax, or worst-case, criterion, the type-I and type-II errors are suprema over $\Nc$ and $\Ac$, respectively. In the universal setting considered here, the test has only access to these sets, but its errors are evaluated at the actual null and alternative states.

Composite quantum hypothesis testing has been studied in \cite{On_Composite_Quantum_Hypothesis_Testing, Composite_two_alternatives}. While the formulation and structure of the competing sets may vary, the optimal error exponent can involve regularized quantities and optimizations over states in the sets, rather than the single-letter expressions that arise in the classical setting. In particular, \cite[Theorem IV.3]{Composite_two_alternatives} shows that, for a single null state and two alternatives, the optimal worst-case error exponent can be strictly below the smaller of the two Umegaki relative entropies from the null to the alternatives.

In this work, we study the one-sample setting of universal hypothesis testing in which the null hypothesis is a known state, and the alternative may be any other quantum state. Thus, the sequence of tests may depend on the null state, but not on the true alternative state. The type-II error exponent is evaluated separately for each alternative.

In the classical setting, Hoeffding \cite{Hoeffding} established a sequence of universal tests that asymptotically achieves, for every alternative, the optimal error exponent of the corresponding simple hypothesis test, namely, the relative entropy between the underlying probability distributions of the null and alternative hypotheses. Since this exponent matches the simple-hypothesis converse, it is universally optimal. In the commuting quantum setting, the classical expression is directly recovered: when the null and alternative states share a common eigenbasis, the classical relative entropy of their eigenvalue distributions coincides with the Umegaki relative entropy between the states. For general quantum states, however, whether one sequence of universal tests can achieve the Umegaki relative entropy for every alternative has remained an open question.

Hoeffding's classical sequence of universal tests exploits the empirical distribution, or type, which records the frequencies of observed outcomes and consistently estimates the underlying probability distribution. The method of types and Sanov's theorem \cite[Theorem 11.4.1]{ElementsOfInformationTheoryBook} control the probabilities of empirical distributions that are atypical under the true state, with decay rates given by the Kullback--Leibler divergence. This large-deviation principle is the key ingredient in Hoeffding's construction and explains how one test can achieve the optimal pointwise exponent for every alternative.
The quantum setting has no direct analogue of the classical empirical distribution that yields the same method-of-types argument. While quantum versions of Sanov's theorem have been developed using different formulations \cite{AQuantumVersionOfSanovsTheorem, Hayashi_Sanov}, the formulation in \cite{AQuantumVersionOfSanovsTheorem} is less general than the classical Sanov theorem. The work in \cite{Hayashi_Sanov} considers a quantum analogue of the empirical distribution, but obtains an exponent based on a divergence that is smaller than the quantum relative entropy. A recent work \cite{Towards_Quantum_Universal_Hypothesis_Testing} studied a universal quantum hypothesis testing framework based on state tomography. The exponent obtained there, which is based on the trace distance between quantum states, is generally weaker than the Umegaki relative-entropy exponent, as demonstrated by the quantum Pinsker inequality \cite{QuantumPinksterInequality}. That work also discusses the quantum Sanov result in \cite{Hayashi_Sanov} and its associated exponent.

Quantum state tomography is considered to be a natural candidate to replace the empirical distribution from the classical setting. Its performance can be described in terms of target fidelity or trace distance, while its large-deviation rate function governs the asymptotic exponential decay of the probability of obtaining an estimate far from the true state. Keyl \cite{QuantumStateEstimationAndLargeDeviations} conjectured the optimal rate function for covariant quantum state tomography. Subsequent work \cite{OptimalRateFunction} identified it as the reverse sandwiched relative entropy, a quantity introduced in \cite[Eq. (9)]{alpha-z-renyi-entropies}, and confirmed Keyl's conjecture. Keyl's algorithm is based on Schur--Weyl methods used in spectrum estimation, entropy estimation, and state tomography, which are described in \cite{EstimatingSpectrum, EstimatingQuantumEntropy, SampleOptimalTomographyOfQuantumStates, Sampple_optimal_and_memory_efficient_quantum_state_tomography}. The reverse sandwiched relative entropy has also appeared as an optimal error exponent in particular composite quantum hypothesis testing problems \cite{Operational_interpretation_of_the_reverse_sandwiched_renyi}. In addition, the work in \cite{A_Quantum_Method_Of_Types} introduces a sequence of universal tests for composite hypothesis testing with an error exponent given by the reverse sandwiched relative entropy between the closest states from the respective hypothesis sets.

Here, we consider the one-sample universal quantum hypothesis testing problem described above. While the quantum Stein lemma provides a sequence of tests achieving the Umegaki relative entropy for each fixed alternative, that sequence may depend on the alternative. We show that, unless the null state is maximally mixed, no single sequence of tests can achieve the Umegaki relative entropy for every alternative simultaneously. We do not impose any structural assumptions, such as covariance, on the tests. We prove this by explicitly constructing, for each given non-maximally mixed null state, a composite hypothesis testing problem with two alternatives for which simultaneous achievability is impossible. Additionally, we construct a sequence of universal tests based on Schur--Weyl duality whose error exponent is at least the reverse sandwiched relative entropy for every alternative state. This generalizes \cite{A_Quantum_Method_Of_Types}, which shows this result for the worst-case performance. In the commuting setting, this exponent agrees with the Umegaki quantum relative entropy. Thus, in this case, the proposed sequence of tests is universally optimal.

The rest of the paper is organized as follows. Section~\ref{sec:setting} presents the setting and background. Section~\ref{sec:main_results} states the main results and compares the relevant relative entropies. Section~\ref{sec:numerical_example} presents a numerical example that illustrates the performance of the proposed sequence of universal tests and compares it to that of the optimal simple test. Section~\ref{sec:tomography} discusses the connection between universal hypothesis testing and quantum state tomography. Section~\ref{sec:converse} proves the converse result, and Section~\ref{sec:achievability} gives the explicit achievability construction. The appendix contains part of the derivation used in the two-dimensional case of the converse.

\subsection{Notation}
Throughout this paper, $\Hc$ denotes a Hilbert space of finite dimension $d$ and $\Sc(\Hc)$ the corresponding set of density operators. We denote the identity operator by $\mathbb{I}$, unitary operators by $U$, the trace by $\operatorname{Tr}[\cdot]$, the trace norm by $\|A\|_1 \coloneqq \operatorname{Tr}[\sqrt{A^\dagger A}]$, the support of an operator by $\operatorname{supp}(\cdot)$, and the adjoint by $(\cdot)^\dagger$. For Hermitian operators $A$ and $B$, we write $A \succeq 0$ if $A$ is positive semidefinite and $A \succeq B$ if $A - B \succeq 0$. For operators $A$ and $B$, we write $A\otimes B$ and $A\oplus B$ for their tensor product and direct sum, respectively, and $\cong$ for unitary equivalence (isomorphism) of Hilbert spaces. The Umegaki relative entropy is denoted by
\begin{align}
    D(\rho\,\|\,\sigma) \coloneqq
    \begin{cases}
        \operatorname{Tr}[\rho\log\rho-\rho\log\sigma], & \text{if } \operatorname{supp}(\rho)\subseteq\operatorname{supp}(\sigma),\\
        +\infty, & \text{otherwise}.
    \end{cases}
\end{align}
The von Neumann entropy is denoted by $H(\rho)=-\operatorname{Tr}[\rho\log\rho]$. For $s \in (0,1)$, the sandwiched Rényi relative entropy of order $s$ is denoted by
\begin{align}
    \widetilde{D}_s(\rho\,\|\,\sigma) \coloneqq \frac{1}{s-1}\log\operatorname{Tr}\left[\left(\sigma^{\frac{1-s}{2s}}\rho\sigma^{\frac{1-s}{2s}}\right)^s\right]
\end{align}
(with $\log 0 \coloneqq -\infty$). Moreover, for $s \in (0,1)$, the reverse sandwiched Rényi relative entropy of order $s$ is denoted by
\begin{align}\label{reverse_relative_entropy}
    D_{R,s}(\rho\,\|\,\sigma) \coloneqq \frac{1}{s-1}\log\operatorname{Tr}\left[\left(\rho^{\frac{s}{2(1-s)}}\sigma\rho^{\frac{s}{2(1-s)}}\right)^{1-s}\right],
\end{align}
and the reverse sandwiched relative entropy is defined as
\begin{align}
    D_R(\rho\,\|\,\sigma) \coloneqq \lim_{s\to1^-} D_{R,s}(\rho\,\|\,\sigma).
\end{align}
Here, $\log$ denotes the natural logarithm.
\section{Setting and Background}\label{sec:setting}
In quantum hypothesis testing (QHT), one is given $n$ independent and identically distributed (i.i.d.) copies of an unknown quantum state. In the binary setting, we have two competing hypotheses, the null hypothesis and the alternative hypothesis. In its simplest form, the objective is to determine whether the underlying state coincides with the null state $\rho$ or with the alternative state $\sigma$. This is usually stated as
\begin{align}\label{simple_errors}
    \begin{cases}
        \mathrm{H}_0: \text{the true state is }\rho,\\
        \mathrm{H}_1: \text{the true state is } \sigma.
    \end{cases}
\end{align}
A test or decision rule on $n$ copies is defined by a two-outcome positive operator-valued measure
(POVM). We denote the POVM by $\{T_n,\mathbb{I}-T_n\}$, where $0\preceq T_n \preceq \mathbb{I}$. The outcome $0$ leads to the acceptance of $\mathrm{H}_0$, while $1$ leads to
its rejection in favor of $\mathrm{H}_1$. Associated with the decision rule, we can have two types of error:
\begin{itemize}
    \item \textbf{Type-I error probability} (incorrectly rejecting $\mathrm{H}_0$ when the true state is $\rho$):
    \begin{align}\label{eq: type_I_error}
        \alpha_n(\rho\mid T_n) \coloneqq \operatorname{Tr}\left[ (\mathbb{I} - T_n) \rho^{\otimes n} \right].
    \end{align}
    \item \textbf{Type-II error probability} (incorrectly accepting $\mathrm{H}_0$ when the true state is $\sigma$):
    \begin{align}\label{eq: type_II_error}
        \beta_n(\sigma\mid T_n) \coloneqq \operatorname{Tr}\left[ T_n \sigma^{\otimes n} \right].
    \end{align}
\end{itemize}
We consider the Stein regime (i.e., the asymmetric setting), where the type-I error probability is bounded by a constant $\epsilon \in (0, 1)$, and we aim to maximize the rate of exponential decay of the type-II error probability as $n \to \infty$. For a fixed sequence of decision rules $\{T_n\}_{n=1}^\infty$ whose type-I error is bounded by $\epsilon$, we define its error exponent by
\begin{align}\label{eq:error_exponent}
    E(\rho \,\|\, \sigma\mid \{T_n\}) \coloneqq \liminf_{n \to \infty} -\frac{1}{n} \log \beta_n(\sigma\mid T_n).
\end{align}
\subsection{The Quantum Stein Lemma}
For each number of copies $n$, let the optimal type-II error probability under a fixed type-I error threshold $\epsilon$ be given by
\begin{align}
    \beta^{\star}_{n,\epsilon}(\rho\,\|\,\sigma) \coloneqq \inf_{0\preceq T_n\preceq\mathbb{I}}\big\{ \beta_n(\sigma\mid T_n):\,  \alpha_n(\rho\mid T_n)\leq \epsilon \big\}.
\end{align}
The best achievable error exponent is then
\begin{align}
    E^\star(\rho\,\|\,\sigma )\coloneqq  \lim_{\epsilon\rightarrow 0^+}\liminf_{n\rightarrow\infty}\left\{-\frac{1}{n} \log \beta^\star_{n,\epsilon}(\rho\,\|\,\sigma)\right\}.
\end{align}
The outer limit imposes a vanishing type-I error threshold. In the simple i.i.d. setting, the quantum Stein lemma implies that the optimal exponent is the same for every fixed $\epsilon\in(0,1)$, so this outer limit does not change the value. Here, we retain it to make the vanishing-error convention explicit and consistent with the composite setting.
In classical hypothesis testing, the Chernoff--Stein lemma \cite[Section 11.8]{ElementsOfInformationTheoryBook} identifies the optimal error exponent for simple hypotheses with the Kullback--Leibler divergence between the two distributions. The quantum Stein lemma establishes that the Umegaki relative entropy is the corresponding optimal exponent for simple quantum hypotheses \cite{ProperFormulaRelativeEntropy, Strong_converse_and_Stein_s_lemma_in_quantum_hypothesis_testing}, i.e.,
\begin{align}
    E^\star (\rho\,\|\,\sigma )=D(\rho \,\|\, \sigma). 
\end{align}
\subsection{Composite Hypothesis Testing}\label{sec:composite_testing}
Composite hypothesis testing concerns binary tests in which one or both hypotheses are sets of states. Given $n$ copies of an unknown state, the test determines whether it belongs to a null set $\Nc\subset\Sc(\Hc)$ or an alternative set $\Ac\subset\Sc(\Hc)$:
\begin{align}
    \begin{cases}
        \mathrm{H}_0: \text{the true state is } \rho \in \Nc,\\
        \mathrm{H}_1: \text{the true state is } \sigma \in \Ac.
    \end{cases}
\end{align}
As in the simple setting, the idea is to construct a test or decision rule on $n$ copies that is defined by a two-outcome positive operator-valued measure (POVM). We denote it by $T_n$, where $0 \preceq T_n \preceq \mathbb{I}$. The outcome $0$ leads to the acceptance of $\mathrm{H}_0$, while $1$ leads to its rejection in favor of $\mathrm{H}_1$. 
In the composite setting, a common performance criterion is the worst-case (or minimax) criterion. In this case, the error probabilities for a test $T_n$ are:
\begin{itemize}
    \item \textbf{Type-I composite error probability} (incorrectly rejecting $\mathrm{H}_0$ when the true state belongs to $\Nc$):
    \begin{align}
        \alpha_n^C(\Nc\mid T_n) \coloneqq \sup_{\rho\in\Nc}\operatorname{Tr}\left[ (\mathbb{I} - T_n) \rho^{\otimes n} \right].
    \end{align}
    \item \textbf{Type-II composite error probability} (incorrectly accepting $\mathrm{H}_0$ when the true state belongs to $\Ac$):
    \begin{align}
        \beta_n^C(\Ac\mid T_n) \coloneqq \sup_{\sigma\in\Ac} \operatorname{Tr}\left[ T_n \sigma^{\otimes n} \right].
    \end{align}
\end{itemize}
We consider again the Stein regime, where the type-I composite error probability is bounded by a constant $\epsilon \in (0, 1)$, and we aim to maximize the rate of exponential decay of the type-II composite error probability as $n \to \infty$. For a fixed sequence of decision rules $\{T_n\}_{n=1}^\infty$ whose worst-case type-I error over the null set is bounded by $\epsilon$, we define its error exponent by
\begin{align}
    E_{C}(\Nc \,\|\, \Ac\mid \{T_n\}) \coloneqq \liminf_{n \to \infty} -\frac{1}{n} \log \beta_n^C(\Ac\mid T_n).
\end{align}
For each number of copies $n$, define the optimal composite type-II error probability under a fixed type-I error threshold $\epsilon$ by
\begin{align}
    \beta^{C,\star}_{n,\epsilon}(\Nc\,\|\,\Ac) \coloneqq \inf_{0\preceq T_n\preceq\mathbb{I}}\{ \beta_n^C(\Ac\mid T_n):\,  \alpha_n^C(\Nc\mid T_n)\leq \epsilon \}.
\end{align}
The best achievable composite error exponent is then
\begin{align}
    E_C^\star(\Nc\,\|\,\Ac )\coloneqq  \lim_{\epsilon\rightarrow 0^+}\liminf_{n\rightarrow\infty}\left\{-\frac{1}{n} \log \beta^{C,\star}_{n,\epsilon}(\Nc\,\|\,\Ac)\right\}.
\end{align}
Note that we can recover the simple setting by choosing $\Nc = \{\rho\}$ and $\Ac = \{\sigma\}$. 
\subsection{Universal Hypothesis Testing and Hoeffding's Result}
Universal quantum hypothesis testing (UQHT) is a special case of composite hypothesis testing, where the test can only depend on the sets of states $\Nc$ and $\Ac$, but the error probabilities are not maximized over $\Nc$ and $\Ac$. Thus, the worst-case error probabilities in Section~\ref{sec:composite_testing} are replaced by the error probabilities \eqref{eq: type_I_error} and \eqref{eq: type_II_error}, evaluated at the true states $\rho\in\Nc$ and $\sigma\in\Ac$.

In this paper, we focus on the universal one-sample testing problem, where $\Nc = \{\rho\}$ and $\Ac = \Sc(\Hc)\setminus \{\rho\}$. Thus, the test has access to the null state $\rho$, but the alternative state can be any state other than $\rho$:
\begin{align}
    \begin{cases}
        \mathrm{H}_0: \text{the true state is } \rho,\\
        \mathrm{H}_1: \text{the true state is } \sigma \in \Sc(\Hc)\setminus\{\rho\}.
    \end{cases}
\end{align}
A sequence of universal tests $\{T_n\}$ may depend on the null state $\rho$, but not on the true alternative $\sigma$, while its type-II error exponent is evaluated separately for each alternative, according to \eqref{eq:error_exponent}. 

In the classical setting, Hoeffding proved that there exists a sequence of universal tests that achieves the Chernoff--Stein exponent, i.e., the Kullback--Leibler divergence, for every alternative \cite{Hoeffding}. Since this is also the optimal exponent in the simple setting, the simple-hypothesis converse applies to each alternative, and the Hoeffding test is universally optimal in the sense that it achieves the optimal exponent $D(\rho\,\|\,\sigma)$ for every alternative.
In the quantum setting, whether a single sequence of universal tests $\{T_n\}$ can simultaneously achieve $E(\rho\,\|\,\sigma\mid\{T_n\}) = D(\rho\,\|\,\sigma)$ for all $\sigma \in \Sc(\Hc)\setminus\{\rho\}$ has remained a central open question.
\section{Main Results}\label{sec:main_results}
We now state the two main results of the paper. The first shows that the quantum relative entropy cannot be achieved simultaneously for every alternative in universal quantum hypothesis testing. The second establishes that there exists a sequence of universal tests that achieves the reverse sandwiched Rényi relative entropy.
\begin{theorem} 
    \label{main_result_converse}
    Consider the UQHT setting with null hypothesis $\rho \neq \mathbb{I}/d$. Then, there is no sequence of tests $\{T_n\}$ that achieves the relative entropy $D(\rho\,\|\,\sigma)$ as an error exponent for every alternative state, i.e.,
    \begin{align} 
        \nexists \{T_n\} \quad\text{such that}\quad E(\rho\,\|\,\sigma\mid \{T_n\}) = D(\rho\,\|\,\sigma), \quad\forall \sigma\in \Sc(\Hc)\setminus\{\rho\}.
    \end{align}
\end{theorem}

\begin{proof}
The proof is given in Section~\ref{sec:converse}.
\end{proof}
For each fixed alternative $\sigma$, the quantum Stein lemma guarantees a sequence of tests (which may depend on $\sigma$) with exponent $D(\rho\,\|\,\sigma)$. Theorem~\ref{main_result_converse} rules out a sequence of universal tests achieving these pointwise optima for all alternatives simultaneously. The proof constructs a composite hypothesis testing problem with two alternatives with equal relative entropies for which the optimal worst-case exponent $E_C^\star(\Nc \,\|\,\Ac) $ is strictly smaller than their common value, contradicting the existence of a sequence of universal tests achieving the relative entropy for both alternatives and consequently a sequence of universal tests achieving the relative entropy for all alternatives.
This raises the question of what the appropriate notion of optimality should be for universal quantum hypothesis testing. Indeed, there is a trade-off between requiring a test to perform particularly well for a specific alternative and requiring it to perform reasonably well for a large set of alternative states.

Our second main result demonstrates that there exists a sequence of universal tests whose error exponent in the Stein regime is at least the reverse sandwiched Rényi relative entropy.
\begin{theorem}
    \label{main_result_achievability}
    Consider the UQHT setting. For every fixed $s\in(0,1)$, there exists a sequence of universal tests $\{T_n^{(s)}\}$ such that
    \begin{align}\label{eq:achievable_exponent}
        E(\rho\,\|\,\sigma \mid \{T_n^{(s)}\}) \geq D_{R,s}(\rho\,\|\,\sigma),\quad \forall\sigma\in\Sc(\Hc)\setminus\{\rho\}.
    \end{align}
\end{theorem}
\begin{proof}
The proof is given in Section~\ref{sec:achievability}.
\end{proof}
We next compare the achievable exponent \eqref{eq:achievable_exponent} in the limit as $s\to 1^-$ with the Umegaki quantum relative entropy, which is the optimal exponent in the simple setting. In the derivation we will make use of the Araki--Lieb--Thirring (ALT) inequality:
\begin{lemma}[Araki--Lieb--Thirring inequality \cite{ALT-ineq}]
\label{lem:ALT}
Let $A, B \succeq 0$ and let $q \ge 0$. Then:
\begin{itemize}
    \item For $r \ge 1$:
    \begin{align}
        \operatorname{Tr}\left[\left(B^{1/2} A B^{1/2}\right)^{rq}\right] \le \operatorname{Tr}\left[\left(B^{r/2} A^r B^{r/2}\right)^q\right].
    \end{align}
    \item For $0 < r \le 1$, the inequality is reversed:
    \begin{align}
        \operatorname{Tr}\left[\left(B^{1/2} A B^{1/2}\right)^{rq}\right] \ge \operatorname{Tr}\left[\left(B^{r/2} A^r B^{r/2}\right)^q\right].
    \end{align}
\end{itemize}
Equality holds if, and only if, one of the following three cases occurs: $A$ and $B$ commute, $r = 1$, or $q = 0$.
\end{lemma} 
Let $\rho, \sigma \in \Sc(\Hc)$ with $\operatorname{supp}(\rho) \subseteq \operatorname{supp}(\sigma)$. For $s \in [1/2, 1)$, we invoke the ALT inequality, with $r \coloneqq \frac{1-s}{s} \le 1$ and $q \coloneqq s > 0$, so that $rq = 1-s$. Choosing the positive semidefinite operators
\begin{align}
    A \coloneqq \sigma, \quad \text{and} \quad B \coloneqq \rho^{\frac{s}{1-s}},
\end{align}
the ALT inequality yields
\begin{align}\label{eq:ALT_step}
    \operatorname{Tr}\left[\left(\rho^{\frac{s}{2(1-s)}}\sigma\rho^{\frac{s}{2(1-s)}}\right)^{1-s}\right] \ge \operatorname{Tr}\left[\left(\rho^{1/2}\sigma^{\frac{1-s}{s}}\rho^{1/2}\right)^s\right].
\end{align}
Let $C=\sigma^{\frac{1-s}{2s}}\rho^{1/2}$. Since $C^\dagger C$ and $CC^\dagger$ have the same eigenvalues, we can write \eqref{eq:ALT_step} as
\begin{align}
    \operatorname{Tr}\left[\left(\rho^{\frac{s}{2(1-s)}}\sigma\rho^{\frac{s}{2(1-s)}}\right)^{1-s}\right] \geq \operatorname{Tr}\left[\left(\sigma^{\frac{1-s}{2s}}\rho\sigma^{\frac{1-s}{2s}}\right)^s\right],
\end{align}
with equality when $\rho$ and $\sigma$ commute. Taking the logarithm and dividing by $s - 1$, we then obtain that
\begin{align}
    D_{R,s}(\rho\,\|\,\sigma) & = \frac{1}{s-1}\log\operatorname{Tr}\left[\left(\rho^{\frac{s}{2(1-s)}}\sigma\rho^{\frac{s}{2(1-s)}}\right)^{1-s}\right] \notag 
    \\&\le \frac{1}{s-1}\log\operatorname{Tr}\left[\left(\sigma^{\frac{1-s}{2s}}\rho\sigma^{\frac{1-s}{2s}}\right)^s\right] \notag \\
    &= \widetilde{D}_s(\rho\,\|\,\sigma).
\end{align}
Recall that $\widetilde{D}_s(\rho\,\|\,\sigma)$ denotes the sandwiched Rényi relative entropy of order $s$. Here, the inequality is reversed because $s-1<0$. For $s\ge 1/2$, the sandwiched Rényi relative entropy is monotonically increasing in its order and converges to the Umegaki relative entropy as $s$ tends to one~\cite{MullerLennert2013, Wilde2014, Beigi2013}. Combining these facts gives
\begin{align*}
    D_{R}(\rho\,\|\,\sigma) = \lim_{s\to1^-}{D}_{R,s}(\rho\,\|\,\sigma) \le \lim_{s\to1^-}\widetilde{D}_s(\rho\,\|\,\sigma)=D(\rho\,\|\,\sigma),
\end{align*}
with equality if, and only if, $\rho$ and $\sigma$ commute. Thus, the sequence of universal tests discussed in Theorem~\ref{main_result_achievability} achieves $D(\rho\,\|\,\sigma)$ if $\rho$ and $\sigma$ commute, but achieves an exponent strictly smaller than $D(\rho\,\|\,\sigma)$ otherwise. Note however that this does not establish an upper bound on the exponent of the proposed tests. Moreover, fixed preprocessing independent of the alternative can strictly improve the achievable bound, because the reverse sandwiched divergences do not generally satisfy data processing for $s>1/2$.
\section{Numerical Example: Discriminating Qubit States}\label{sec:numerical_example}
To illustrate the results, consider qubit states in the $XZ$-plane of the Bloch sphere. Let $X,Y,Z$ denote the standard Pauli matrices, and fix the null state as
\begin{align}
    \rho = \frac{1}{2}\left(\mathbb{I} + r_\rho Z\right),
\end{align}
with Bloch-vector radius $r_\rho = 0.8$, pointing along the positive $Z$-axis. Let the alternatives have Bloch-vector radius $r_\sigma = 0.81$ and angle $\theta \in [0,\pi]$ relative to $\rho$:
\begin{align}
    \sigma(\theta) = \frac{1}{2}\left(\mathbb{I} + r_\sigma (\sin\theta\,X + \cos\theta\,Z)\right).
\end{align}
The states commute at $\theta\in\{0,\pi\}$ and do not commute for $\theta\in(0,\pi)$.

\begin{figure}[!ht]
    \centering
    \includegraphics[width=0.75\linewidth]{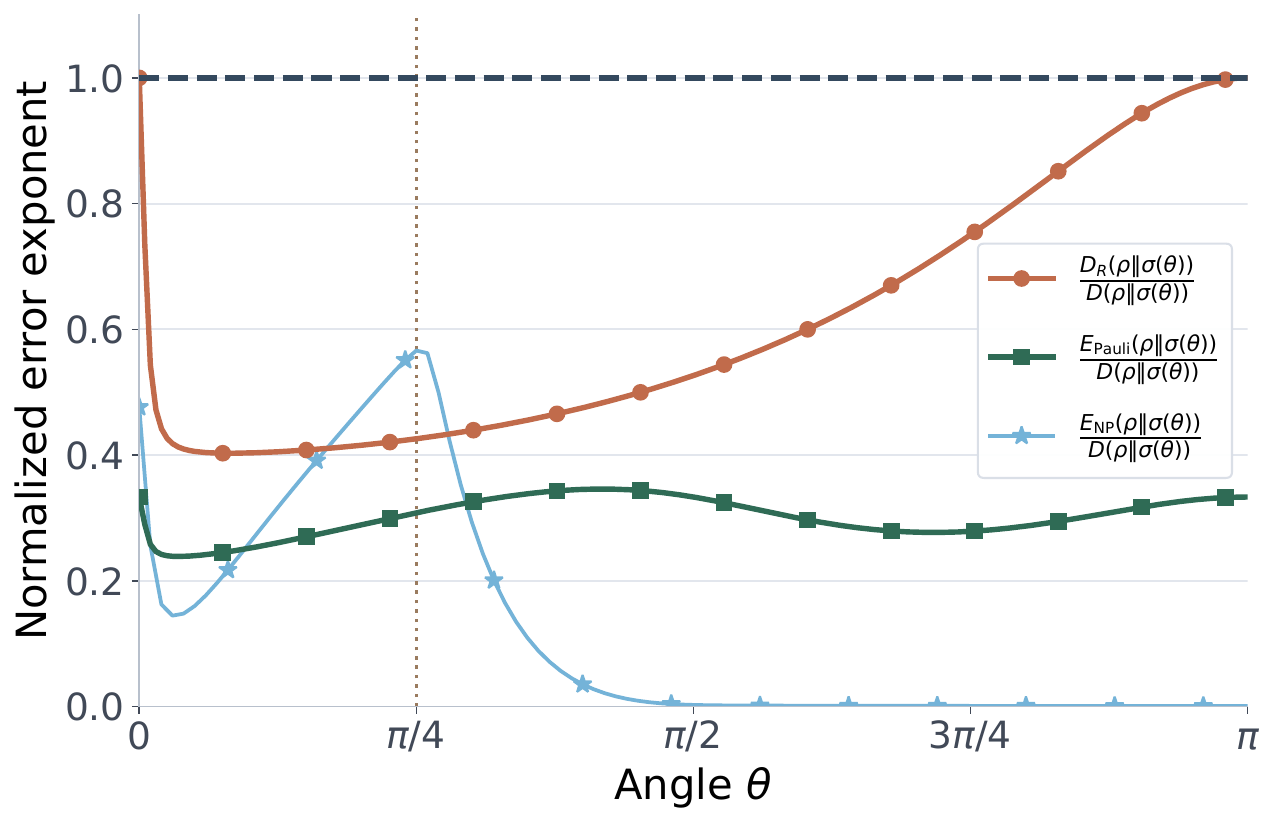}
    \caption{Error exponents normalized by $D(\rho\,\|\,\sigma(\theta))$, for Bloch-vector radii $r_\rho = 0.8$ and $r_\sigma = 0.81$. The curves show the reverse sandwiched relative entropy $D_R(\rho\,\|\,\sigma(\theta))$, the universal Pauli-measurement exponent $E_{\mathrm{Pauli}}(\rho\,\|\,\sigma(\theta))$, and the finite-blocklength exponent $E_{\mathrm{NP}}(\rho\,\|\,\sigma(\theta))$ of the Neyman--Pearson test optimized for $\sigma(\pi/4)$ and held fixed as $\theta$ varies ($n = 128$, $\alpha = 0.01$).}
    \label{fig:numerical_comparison}
\end{figure}

Figure~\ref{fig:numerical_comparison} compares $D_R(\rho\,\|\,\sigma(\theta))$, the universal exponent obtained from single-copy Pauli measurements, and the finite-blocklength exponent of a Neyman--Pearson test. Specifically, measuring an equal fraction of the available copies projectively along each of the three Pauli axes yields the universal error exponent
\begin{align}
    E_{\mathrm{Pauli}}(\rho\,\|\,\sigma(\theta)) = \frac{1}{3}\Big[ D\big(\rho_x\,\|\,\sigma_x(\theta)\big) + D\big(\rho_y\,\|\,\sigma_y(\theta)\big) + D\big(\rho_z\,\|\,\sigma_z(\theta)\big) \Big],
\end{align}
where each subscript $j\in\{x,y,z\}$ denotes the projection of the state onto the eigenbasis of the corresponding Pauli axis (or, equivalently, the binary probability distribution induced by a projective measurement along that axis); see \cite{FerrezueloOteo2026} for details. We also consider the Neyman--Pearson test $T^{\mathrm{NP}}_n$ optimized for the fixed target alternative $\sigma(\pi/4)$ with type-I error threshold $\alpha=0.01$ and $n=128$ copies. We keep this test fixed as $\theta$ varies and evaluate its type-II error for each alternative. The exponent is then defined as
\begin{align}
    E_{\mathrm{NP}}(\rho\,\|\,\sigma(\theta)) \coloneqq -\frac{1}{n}\log \beta_n(\sigma(\theta)\mid T^{\mathrm{NP}}_n).
\end{align}

For all angles $\theta\in(0,\pi)$, $D_R(\rho\,\|\,\sigma(\theta))$ lies strictly above $E_{\mathrm{Pauli}}(\rho\,\|\,\sigma(\theta))$, which remains close to one third of $D(\rho\,\|\,\sigma(\theta))$ in the non-commuting regime. This gap reflects the trade-off between measurement complexity and achievable error exponents: while $E_{\mathrm{Pauli}}$ only requires local, single-copy projective measurements, approaching $D_R(\rho\,\|\,\sigma(\theta))$ via the Schur--Weyl construction requires collective measurements across all $n$ copies. Furthermore, $D_R(\rho\,\|\,\sigma(\theta))=D(\rho\,\|\,\sigma(\theta))$ for $\theta=0$ and $\theta=\pi$, since in these cases $\rho$ and $\sigma(\theta)$ commute.

On the other hand, the Neyman--Pearson curve exceeds $D_R(\rho\,\|\,\sigma(\theta))$ near $\theta=\pi/4$. Since this test is optimized for $\sigma(\pi/4)$, its exponent at that point converges to $D(\rho\,\|\,\sigma(\pi/4))$ as $n$ increases. However, because the test is fixed for a single alternative, its exponent drops quickly for other values of $\theta$. This illustrates the consequences of Theorem~\ref{main_result_converse}, which rules out a sequence of universal tests achieving $D(\rho\,\|\,\sigma(\theta))$ for every alternative simultaneously.
\section{Quantum State Tomography and Universal Tests}\label{sec:tomography}

In the classical setting, Hoeffding's sequence of universal tests is built from the empirical distribution of the samples. Sanov's theorem describes the exponential decay of the probability of observing an atypical empirical distribution, with a rate given by the relative entropy~\cite[Theorem 11.4.1]{ElementsOfInformationTheoryBook}. Quantum state tomography plays an analogous role: it estimates an unknown state from multiple copies and assigns a probability distribution to the possible estimates~\cite{QuantumStateEstimationAndLargeDeviations}.

Formally, a quantum state tomography algorithm is a sequence of POVMs $\{M_n\}_{n=1}^{\infty}$ on the state space $\Sc(\Hc)$, with effects acting on $\Hc^{\otimes n}$. Applied to $n$ copies of an unknown state $\sigma$, its outcome is an estimate $\hat{\rho}\in\Sc(\Hc)$. For a Borel set $\Delta\subseteq\Sc(\Hc)$, write
\begin{equation}
    M_n(\Delta) \coloneqq \int_{\hat{\rho}\in\Delta} M_n(\mathrm{d}\hat{\rho}).
\end{equation}
Clearly, $0\preceq M_n(\Delta)\preceq\mathbb{I}$. Suppose that the POVM can be written with respect to an outcome measure $\mu_n$ as
\begin{equation}
    M_n(\mathrm{d}\hat{\rho}) = M_n(\hat{\rho})\,\mathrm{d}\mu_n(\hat{\rho}),
\end{equation}
where we call $M_n(\hat{\rho})$ a \emph{POVM density} with respect to $\mu_n$. With this representation, the Born rule gives the probability density of obtaining the estimate $\hat{\rho}$ with respect to $\mu_n$ when the true state is $\sigma$:
\begin{equation}
    p_n(\hat{\rho}\mid\sigma)
    \coloneqq \operatorname{Tr}[M_n(\hat{\rho})\sigma^{\otimes n}].
\end{equation}
The probability of an event $\Delta$ is then
\begin{equation}
    P_n(\Delta\mid\sigma)
    \coloneqq \mathbb{P}_{\sigma}(\hat{\rho}\in\Delta)
    = \int_{\hat{\rho}\in\Delta} p_n(\hat{\rho}\mid\sigma)\,\mathrm{d}\mu_n(\hat{\rho})
    = \operatorname{Tr}[M_n(\Delta)\sigma^{\otimes n}].
\end{equation}
The algorithm is said to be \emph{consistent} if its estimate converges in probability to the true state, i.e., for every $\rho\in\Sc(\Hc)$ and every open neighborhood $\Delta(\rho)$ of $\rho$,
\begin{equation}\label{eq:tomography_consistency}
    \lim_{n\to\infty}\operatorname{Tr}[M_n(\Delta(\rho))\rho^{\otimes n}] = 1.
\end{equation}
This is equivalent to saying that, for every $\epsilon >0$, there exists an integer $N$ such that, for all $n\geq N$,
\begin{equation}
    \operatorname{Tr}[M_n(\Delta(\rho))\rho^{\otimes n}] \geq 1-\epsilon.
\end{equation}
Consistency ensures that the estimate converges to the true state, but it does not specify how quickly the probability of an atypical estimate decays. This asymptotic behavior is encapsulated in the notion of a rate function. We recall the definition from \cite[Definitions~A.1 and A.2]{QuantumStateEstimationAndLargeDeviations}:

\begin{definition}[Rate function \cite{QuantumStateEstimationAndLargeDeviations}]\label{def:rate_function}
Fix a true state $\sigma\in\Sc(\Hc)$. A lower semicontinuous function $I(\cdot\,\|\,\sigma):\Sc(\Hc)\to[0,+\infty]$ with $I(\hat{\rho}\,\|\,\sigma)=0$ if, and only if, $\hat{\rho}=\sigma$ is a rate function for the tomography algorithm $\{M_n\}_{n=1}^\infty$ if, for every Borel set $\Delta\subseteq\Sc(\Hc)$,
\begin{equation}\label{eq:ldp_bounds}
    \inf_{\hat{\rho}\in\overline{\Delta}}I(\hat{\rho}\,\|\,\sigma)
    \le \liminf_{n\to\infty}-\frac{1}{n}\log P_n(\Delta\mid\sigma)
    \le \limsup_{n\to\infty}-\frac{1}{n}\log P_n(\Delta\mid\sigma)
    \le \inf_{\hat{\rho}\in\Delta^\circ}I(\hat{\rho}\,\|\,\sigma),
\end{equation}
where $\Delta^\circ$ and $\overline{\Delta}$ denote the interior and the closure of $\Delta$ in $\Sc(\Hc)$, respectively. In particular, whenever
\begin{equation*}
    \inf_{\hat{\rho}\in\Delta^\circ}I(\hat{\rho}\,\|\,\sigma)
    =\inf_{\hat{\rho}\in\overline{\Delta}}I(\hat{\rho}\,\|\,\sigma),
\end{equation*}
the limit as $n\to\infty$ exists and is given by
\begin{equation}\label{eq:rate_limit}
    \lim_{n\to\infty}-\frac{1}{n}\log P_n(\Delta\mid\sigma)
    =\inf_{\hat{\rho}\in\Delta}I(\hat{\rho}\,\|\,\sigma).
\end{equation}
\end{definition}

We next show that for every quantum state tomography algorithm with a continuous rate function $I$, we can find a sequence of universal tests whose error exponent is lower-bounded by $I(\rho\,\|\,\sigma)$.

\begin{theorem}\label{thm:tomography_to_quht}
Let $\{M_n\}_{n=1}^{\infty}$ be a consistent quantum state tomography algorithm with a continuous rate function $I$. For any fixed null state $\rho$, there is a sequence of universal tests $\{T_n\}_{n=1}^{\infty}$ such that
\begin{equation}
    E(\rho\,\|\,\sigma\mid\{T_n\}) \geq I(\rho\,\|\,\sigma),
    \qquad \forall \sigma\in\Sc(\Hc)\setminus\{\rho\}.
\end{equation}
\end{theorem}

\begin{proof}
We build the test from the tomography outcome. For every number of copies $n$, it accepts the null hypothesis when the estimate is sufficiently close to $\rho$, with closeness measured using the closed trace-norm ball
\begin{equation}
    \Bc_{\delta}(\rho)
    \coloneqq \left\{\hat{\rho}\in\Sc(\Hc):
    \|\hat{\rho}-\rho\|_1\leq\delta\right\}.
\end{equation}
Every such ball contains an open neighborhood of its center. Therefore, consistency implies that, for every integer $k\geq2$, there is a threshold $N_k'$ such that
\begin{equation}\label{eq:coupled_type1}
    \operatorname{Tr}[M_n(\Bc_{1/k}(\rho))\rho^{\otimes n}]
    \geq 1-\frac{1}{k}, \qquad n\geq N_k'.
\end{equation}
Set $N_1=1$, and recursively choose $N_k\geq N_k'$ with $N_k>N_{k-1}$ for $k\geq2$. Further set $\delta_n=1/k$ whenever $N_k\leq n<N_{k+1}$. The radii depend only on the null state, not on the alternative, so this defines a sequence of universal tests
\begin{equation}
    T_n \coloneqq M_n(\Bc_{\delta_n}(\rho)).
\end{equation}
For $N_k\leq n<N_{k+1}$ with $k\geq2$, the type-I error is
\begin{align}
    \alpha_n(\rho\mid T_n)
    = 1-\operatorname{Tr}[T_n\rho^{\otimes n}] \leq \frac{1}{k},
\end{align}
by \eqref{eq:coupled_type1}. Since $k\to\infty$ with $n$, the type-I error vanishes.

Now, fix any alternative state $\sigma\in\Sc(\Hc)\setminus\{\rho\}$ and any radius $\delta>0$, independent of $n$. Since $\delta_n\to0$, we have, for $n$ sufficiently large, $\Bc_{\delta_n}(\rho)\subseteq\Bc_{\delta}(\rho)$. Also, for any $\hat{\rho}\in\Bc_{\delta}(\rho)$, the state $(1-t)\rho+t\hat{\rho}$ lies in the interior of $\Bc_{\delta}(\rho)$ for $0\leq t<1$ and converges to $\hat{\rho}$ as $t\to1^-$. Therefore, the interior of $\Bc_{\delta}(\rho)$ is dense in the closed ball. Continuity of $I(\cdot\,\|\,\sigma)$ makes the infima over the interior and the closed ball equal, so \eqref{eq:rate_limit} applies. Since
\[
    \beta_n(\sigma\mid T_n)
    =P_n(\Bc_{\delta_n}(\rho)\mid\sigma)
    \leq P_n(\Bc_{\delta}(\rho)\mid\sigma)
\]
for sufficiently large $n$, we have
\begin{align}
    E(\rho\,\|\,\sigma\mid\{T_n\})
    &= \liminf_{n\to\infty}-\frac{1}{n}\log\beta_n(\sigma\mid T_n)\notag\\
    &\geq \lim_{n\to\infty}-\frac{1}{n}\log
    P_n\bigl(\Bc_{\delta}(\rho)\mid\sigma\bigr)\notag\\
    &= \inf_{\hat{\rho}\in\Bc_{\delta}(\rho)}
    I(\hat{\rho}\,\|\,\sigma).
\end{align}
This is true for every $\delta> 0$, so we can let $\delta$ tend to zero from above. The continuity of $I(\cdot\,\|\,\sigma)$ then implies that
\begin{align}
    \lim_{\delta\to0^+}\inf_{\hat{\rho}\in\Bc_{\delta}(\rho)} I(\hat{\rho}\,\|\,\sigma) = I(\rho\,\|\,\sigma).
\end{align}
Thus, the same sequence of tests has an exponent greater than or equal to $I(\rho\,\|\,\sigma)$ for every alternative, as claimed.
\end{proof}

Theorem~\ref{thm:tomography_to_quht} is in the spirit of Lemma~5.3 in \cite{QuantumStateEstimationAndLargeDeviations}, which constructs a test for simple hypothesis testing using a tomography algorithm. The acceptance regions are centered only at the fixed null state and are chosen independently of the alternative, as in the classical Hoeffding test. This therefore yields a sequence of universal tests that achieves an exponent of at least $I(\rho\,\|\,\sigma)$ for every alternative $\sigma$.

For covariant quantum state tomography, Keyl~\cite{QuantumStateEstimationAndLargeDeviations} conjectured the optimal large-deviation rate function, and recent work~\cite{OptimalRateFunction} proved that it is given by the reverse sandwiched relative entropy $I(\hat{\rho}\,\|\,\sigma) = D_R(\hat{\rho}\,\|\,\sigma)$. Since $D_R(\cdot\,\|\,\sigma)$ is continuous on full-rank states, applying Theorem~\ref{thm:tomography_to_quht} to the covariant tomography protocol of~\cite{OptimalRateFunction} directly recovers the asymptotic error exponent $D_R(\rho\,\|\,\sigma)$ obtained in the limit $s\to 1^-$ of Theorem~\ref{main_result_achievability}. Notice, however, that Theorem~\ref{thm:tomography_to_quht} relies on a general procedure for constructing a sequence of tests from an estimation scheme, whereas Theorem~\ref{main_result_achievability} provides an explicit, direct sequence of universal tests for UQHT based on Schur--Weyl duality and tilted pretty-good measurements, yielding achievable lower bounds $D_{R,s}(\rho\,\|\,\sigma)$ for every fixed $s\in(0,1)$.

\section{Proof of Theorem~\ref{main_result_converse}}\label{sec:converse}
To prove the theorem, we study the composite problem with null set $\Nc=\{\rho\}$ and alternative set $\Ac=\{\sigma_1,\sigma_2\}$. We use the definitions in Section~\ref{sec:composite_testing}, specialized to these sets (where we write $\rho$ for $\{\rho\}$). In particular,
\begin{align}
    \beta_n^C(\{\sigma_1,\sigma_2\}\mid T_n)
    =\max\{\beta_n(\sigma_1\mid T_n),\,\beta_n(\sigma_2\mid T_n)\}.
\end{align}
We show that there exist alternative states for which there is a strict gap between the optimal composite exponent and the minimum of the two corresponding quantum relative entropies. This is in contrast to the case of commuting states, where
\begin{align}
    E_C^\star(\rho\,\|\,\{\sigma_1,\, \sigma_2\})=    \min_{\sigma\in\{\sigma_1,\sigma_2\}}D(\rho\,\|\,\sigma).
\end{align}
It is worth emphasizing that a composite exponent guarantee is inherently weaker than a simultaneous pointwise exponent guarantee. In a composite test, one only ensures that the worst-case error exponent matches $\min \{ D(\rho \,\|\, \sigma_1), D(\rho \,\|\, \sigma_2) \}$. This guarantees performance at least as good as the minimum relative entropy across all states in the set, but it does not guarantee that the test simultaneously achieves the individual pointwise exponent $D(\rho \,\|\, \sigma_i)$ for the state with the larger relative entropy.
Our results show that even this weaker worst-case guarantee fails to reach $\min \{ D(\rho \,\|\, \sigma_1), D(\rho \,\|\, \sigma_2) \}$ in the non-commuting case.

Our proof builds on \cite{Composite_two_alternatives}, which constructs a composite quantum hypothesis testing problem with two alternatives that fails to achieve the classically optimal exponent $\min_{i=1,2} D(\rho \,\|\, \sigma_i)$. We extend this result by proving that for every null state $\rho \in \Sc(\Hc)$ distinct from the maximally mixed state, there exists a pair of alternative states $\{\sigma_1, \sigma_2\}$ for which the classical optimal exponent cannot be achieved.

To establish this general non-achievability result, we follow the approach of \cite{Composite_two_alternatives} and reduce the composite hypothesis testing problem with two alternatives to a simple hypothesis testing problem against an effective operator defined via the Kubo--Ando geometric mean $\sigma_1 \# \sigma_2$.
Recall that the geometric mean of two positive definite operators is defined as
\begin{align}
    A\#B \coloneqq B^{1/2}(B^{-1/2}AB^{-1/2})^{1/2}B^{1/2}.
\end{align}
The geometric mean has the following properties.
\begin{proposition}[Properties of the geometric mean~\cite{Composite_two_alternatives}]\label{prop:geometric_mean_properties}
\leavevmode
\begin{enumerate}[label=(\roman*),ref=(\roman*)]
    \item\label{prop:gm_symmetric} For positive definite operators $A$ and $B$,
    \begin{align}
        A\#B = B\#A.
    \end{align}
    \item\label{prop:gm_geometric_arithmetic} For positive definite operators $A$ and $B$,
    \begin{align}
        A\#B \preceq \frac{1}{2}(A+B).
    \end{align}
    \item\label{prop:gm_tensor_product} For positive definite operators $A_1,A_2,B_1,B_2$,
    \begin{align}
        (A_1\otimes A_2)\#(B_1\otimes B_2)
        =(A_1\#B_1)\otimes(A_2\#B_2).
    \end{align}
    \item\label{prop:gm_unitary_covariance} For positive definite operators $A,B$ and any unitary operator $U$,
    \begin{align}
        U(A\#B)U^\dagger=(UAU^\dagger)\#(UBU^\dagger).
    \end{align}
    \item\label{prop:gm_relative_entropy} For density operators $\rho,\sigma_1,\sigma_2$,
    \begin{align}
        D(\rho\,\|\,\sigma_1\#\sigma_2)\geq 0,
        \qquad
        D(\rho\,\|\,\sigma_1\#\sigma_2)=0 \iff \rho=\sigma_1=\sigma_2.
    \end{align}
\end{enumerate}
\end{proposition}
The next lemma relates the error exponent of the composite problem to the Umegaki relative entropy between $\rho$ and $\sigma_1\#\sigma_2$.
\begin{lemma}[{\cite[Lemma~IV.1]{Composite_two_alternatives}}] \label{lemma1}
    For any density operators $\rho,\sigma_1,\sigma_2 \in \Sc(\Hc)$ we have
    \begin{align}
        E_C^\star(\rho\,\|\,\{\sigma_1,\,\sigma_2\}) \leq D(\rho\,\|\,\sigma_1\#\sigma_2).
    \end{align}
\end{lemma}
\begin{proof}
See \cite[Lemma~IV.1]{Composite_two_alternatives}.
\end{proof}
Before the next lemma, consider the following quantity for positive definite operators $A$ and $B$:
\begin{align}\label{Lambda(A,B) definition}
    \Lambda(A,B)\coloneqq\log(A\#B)-\frac{\log A+\log B}{2}.
\end{align}
This quantity relates the logarithm of the geometric mean to the logarithmic average and will be used below to characterize when $A$ and $B$ commute. Indeed, for positive definite operators $A$ and $B$,
\begin{align}
    \Lambda(A,B) = 0 &\iff AB = BA, \label{eq:lambda_commute}\\
    \operatorname{Tr}[\Lambda(A,B)] &= 0. \label{eq:lambda_trace}
\end{align}
The equivalence in \eqref{eq:lambda_commute} follows from the Araki--Lieb--Thirring inequality (see \cite[Lemma~II.13]{Composite_two_alternatives}). The trace identity follows because $\operatorname{Tr}[\log X]=\log\det X$ and from the multiplicativity of the determinant.
\begin{lemma}\label{lemma2}
    Let $\rho\in\Sc(\Hc)$ with $\rho\neq\mathbb{I}/d$. There exists a pair of non-commuting density operators $\sigma_1,\sigma_2 \in \Sc(\Hc)$ such that
    \begin{align}
        D(\rho\,\|\,\sigma_1) = D(\rho\,\|\,\sigma_2)
    \end{align}
    and
    \begin{align}
        \operatorname{Tr}[\rho\Lambda(\sigma_1,\sigma_2)]>0.
    \end{align}
    Conversely, for $\rho=\mathbb{I}/d$, no such pair exists.
\end{lemma}
\begin{proof}
Let 
\begin{align}
    R= \begin{pmatrix}
        r_+ & 0 \\
        0& r_-
    \end{pmatrix},
\end{align}
where $r_+>r_-$ are the highest and lowest eigenvalues of $\rho$.
   Consider the following pair of quantum states 
   \begin{align}
       S_1=\frac{1}{2} \begin{pmatrix}
1+z & x-iy\\
x+iy & 1-z
\end{pmatrix}, &&   S_2= \frac{1}{2} \begin{pmatrix}
1+z & -(x-iy)\\
-(x+iy) & 1-z
\end{pmatrix},
   \end{align}
   where $x,y,z$ are the coordinates of a Bloch vector.
   Note that $S_1$ is just a generic qubit state and $S_2$ is the same state but with a change of sign in the off-diagonal terms. If
   \begin{align}
       Z=\begin{pmatrix}
           1 & 0\\
            0 & -1
       \end{pmatrix},
   \end{align}
   then $S_2=ZS_1Z$. Since $R$ is diagonal, $ZRZ = R$ and unitary invariance of relative entropy gives
   \begin{align}
       D(R\,\|\,S_1) = D(ZRZ\,\|\,ZS_1Z) = D(R\,\|\,S_2).
   \end{align}
   We now show that $\Lambda(S_1,S_2)$ is diagonal. By item~\ref{prop:gm_unitary_covariance} of Proposition~\ref{prop:geometric_mean_properties},
   \begin{align}
       S_1\#S_2 = (ZS_2Z)\#(ZS_1Z) = Z(S_2\#S_1)Z.
   \end{align}
   By item~\ref{prop:gm_symmetric} of Proposition~\ref{prop:geometric_mean_properties}, $S_2\#S_1 = S_1\#S_2$ and 
   \begin{align}\label{diagonalS1S2}
       S_1\#S_2 = Z(S_1\#S_2)Z.
   \end{align}
   Therefore, using \eqref{diagonalS1S2} together with the unitary covariance of the logarithm (which follows from the same unitary invariance underlying item~\ref{prop:gm_unitary_covariance}), the definition in \eqref{Lambda(A,B) definition} gives
   \begin{align}
       Z\Lambda(S_1,S_2)Z = \Lambda(S_1,S_2),
   \end{align}
   which implies that $\Lambda(S_1,S_2)$ is diagonal. Moreover, \eqref{eq:lambda_trace} gives $\operatorname{Tr}[\Lambda(S_1,S_2)]=0$, so $\Lambda(S_1,S_2)$ has the form
   \begin{align}
       \Lambda(S_1,S_2) = \gamma(x,y,z)\begin{pmatrix}
           1 & 0 \\0 &-1
       \end{pmatrix}, \label{coefficient_lambda_xyz}
   \end{align}
   where, as derived in Appendix~\ref{derivation_appendix},
   \begin{align}
       \gamma(x, y ,z) = \arctanh{(z)}-\frac{z}{t}\arctanh{(t)},\quad t=\sqrt{x^2+y^2+z^2}.
   \end{align}
   We can now explicitly compute 
   \begin{align}
       \operatorname{Tr}[R\Lambda(S_1,S_2)] = \gamma(x,y,z)(r_+-r_-).
   \end{align}
   Fix a Bloch vector $(x,y,z)$ satisfying
   \begin{align}
       z<0,\qquad \sqrt{x^2+y^2}\neq0,\qquad t=\sqrt{x^2+y^2+z^2}<1.
   \end{align}
   Then $0<|z|<t<1$. Since $u\mapsto\arctanh(u)/u$ is strictly increasing on $(0,1)$,
   \[
       \frac{\arctanh(|z|)}{|z|}<\frac{\arctanh(t)}{t}.
   \]
   Using that $\arctanh$ is odd and multiplying by $z<0$ gives
   \[
       \arctanh(z)>\frac{z}{t}\arctanh(t),
   \]
   so $\gamma(x,y,z)>0$. The condition $t<1$ also ensures that $S_1$ and $S_2$ are full rank.

This two-dimensional construction can be used to give a sharp result for an arbitrary finite-dimensional state $\rho$.
Assume $\rho\neq \mathbb{I}/d$. Let $r_+>r_-$ be the largest and smallest eigenvalues of 
$\rho$, and choose the corresponding unit eigenvectors $\ket{e_+}, \ket{e_-}$. Let $V: \mathbb{C}^2 \rightarrow \mathrm{span}\{\ket{e_+},\ket{e_-}\}$ be the isometry defined by 
\begin{align}
    V\ket{0} = \ket{e_+},\quad V\ket{1}= \ket{e_-}.
\end{align}
If $d=2$, set 
\begin{align}
    \sigma_1 = VS_1V^\dagger,\quad \sigma_2=VS_2V^\dagger.
\end{align}
Then $V^\dagger\rho V = R$ and unitary invariance of relative entropy gives
\begin{align}
    D(\rho\,\|\,\sigma_j) = D(R\,\|\,S_j),\quad j=1,2.
\end{align}
Hence $D(\rho\,\|\,\sigma_1) = D(\rho\,\|\,\sigma_2)$ by the two-dimensional construction. By item~\ref{prop:gm_unitary_covariance} of Proposition~\ref{prop:geometric_mean_properties} and the unitary covariance of the logarithm,
\begin{align}
    \Lambda(\sigma_1,\sigma_2) = V\Lambda(S_1,S_2)V^\dagger,
\end{align}
and hence
\begin{align}
    \operatorname{Tr}[\rho\Lambda(\sigma_1,\sigma_2)] = \gamma(x,y,z)(r_+-r_-)>0.
\end{align}
If $d>2$, let
\begin{align}
    \Kc = \operatorname{span}\{\ket{e_+},\ket{e_-}\}^{\perp},
\end{align}
choose a full-rank density operator $\tau$ on $\Kc$, and fix $0<p<1$. Define
\begin{align}
    \sigma_j=pVS_jV^\dagger \oplus (1-p)\tau,\quad j = 1,2.
\end{align}
These are full-rank density operators. The common block $\tau$ contributes zero to $\Lambda$, while the scalar factor $p$ cancels on the block $VS_jV^\dagger$. Thus
\begin{align}
    \Lambda(\sigma_1,\sigma_2)=V\Lambda(S_1,S_2)V^\dagger \oplus0_{\Kc},
\end{align}
and again
\begin{align}
    \operatorname{Tr}[\rho\Lambda(\sigma_1,\sigma_2)]=\gamma(x,y,z) (r_+-r_-)>0.
\end{align}
Moreover, since $\rho$ leaves $\mathrm{span}\{\ket{e_+},\ket{e_-}\}$ and $\Kc$ invariant, the two relative entropies satisfy
\begin{align}
    D(\rho\,\|\,\sigma_1)-D(\rho\,\|\,\sigma_2) &= \operatorname{Tr}[\rho(\log\sigma_2-\log\sigma_1)]\notag\\
    &=\operatorname{Tr}[R(\log S_2-\log S_1)]\notag\\
    &=0.
\end{align}
Conversely, if $\rho =\mathbb{I}/d$, then for every positive definite pair $\sigma_1,\sigma_2$,
\begin{align}
    \operatorname{Tr}[\rho\Lambda(\sigma_1,\sigma_2)] = \frac{1}{d}\operatorname{Tr}[\Lambda(\sigma_1,\sigma_2)]=0.
\end{align}
Therefore, strict positivity is impossible for the maximally mixed state.
\end{proof}
\begin{remark}
Lemma~\ref{lemma2} is similar to \cite[Lemma IV.2]{Composite_two_alternatives}, but strengthens it in two directions. First, \cite[Lemma IV.2]{Composite_two_alternatives} establishes the existence of three states $\rho, \sigma_1, \sigma_2$ satisfying
\[
    \min_{i=1,2} D(\rho\,\|\,\sigma_i) > D(\rho\,\|\,\sigma_1\#\sigma_2).
\]
By contrast, Lemma~\ref{lemma2} explicitly constructs the pair of alternatives $\sigma_1, \sigma_2$ for any given null state $\rho \neq \mathbb{I}/d$. Second, while the construction in \cite[Lemma IV.2]{Composite_two_alternatives} requires a Hilbert space of dimension $d \ge 4$, our derivation in Lemma~\ref{lemma2} holds for any dimension $d \ge 2$.
\end{remark}

We are now ready to prove Theorem~\ref{main_result_converse}. By Lemma~\ref{lemma2}, for any density operator $\rho\neq\mathbb{I}/d$, there exists a pair of non-commuting density operators $\sigma_1,\sigma_2 \in \Sc(\Hc)$ such that
    \begin{align}
        D(\rho\,\|\,\sigma_1) = D(\rho\,\|\,\sigma_2)=D_0
    \end{align} 
    and
    \begin{align}
        \min_{i=1,2} D(\rho\,\|\,\sigma_i)- E_C^\star(\rho\,\|\,\{\sigma_1,\sigma_2\}) &= D_0-E_C^\star(\rho\,\|\,\{\sigma_1,\sigma_2\})\notag\\
        &\geq D_0-D(\rho\,\|\,\sigma_1\#\sigma_2)\notag\\
        &= \operatorname{Tr}[\rho\Lambda(\sigma_1,\sigma_2)]>0. \label{positive gap}
    \end{align}
The first equality uses $D(\rho\,\|\,\sigma_1)=D(\rho\,\|\,\sigma_2)=D_0$; the subsequent inequality follows from Lemma~\ref{lemma1}; the second equality uses $D_0=\frac{1}{2}(D(\rho\,\|\,\sigma_1)+D(\rho\,\|\,\sigma_2))$; and strict positivity follows from Lemma~\ref{lemma2}.
This fact bounds the optimal error exponent for composite quantum hypothesis testing with two alternatives.

The inequality \eqref{positive gap} rules out a sequence of universal tests that achieves the quantum relative entropy as an error exponent for every $\sigma\in\Sc(\Hc)\setminus\{\rho\}$.
Indeed, if such a sequence of universal tests existed, it would also apply to the composite problem with alternatives $\sigma_1$ and $\sigma_2$ satisfying $D(\rho\,\|\,\sigma_1)=D(\rho\,\|\,\sigma_2)=D_0$ and $\operatorname{Tr}[\rho\Lambda(\sigma_1,\sigma_2)]>0$. (By Lemma~\ref{lemma2}, one can find such a pair whenever $\rho\neq\mathbb{I}/d$.) The resulting composite exponent would be $E_C(\rho\,\|\,\{\sigma_1,\sigma_2\}\mid\{T_n\})=\min_{i=1,2}D(\rho\,\|\,\sigma_i)=D_0$, contrary to \eqref{positive gap}, which gives $D_0-E_C^\star(\rho\,\|\,\{\sigma_1,\sigma_2\})>0$. This proves Theorem~\ref{main_result_converse}.

\section{Proof of Theorem~\ref{main_result_achievability}}\label{sec:achievability}
This section presents a Schur--Weyl-based achievability scheme for universal quantum hypothesis testing (UQHT). Schur--Weyl methods have been used in quantum estimation theory (see \cite[Chapters 5 and 6]{Harrow} and \cite{ChurchOfSymmetricSubspace} for an introduction), including state spectrum estimation \cite{EstimatingSpectrum}, entropy estimation \cite{EstimatingQuantumEntropy}, quantum state tomography \cite{SampleOptimalTomographyOfQuantumStates, Sampple_optimal_and_memory_efficient_quantum_state_tomography}, and quantum versions of Sanov's theorem \cite{Hayashi_Sanov}. A Schur--Weyl-based tomography protocol was also used to establish the optimal rate function in covariant quantum state tomography \cite{OptimalRateFunction}, given by the reverse sandwiched relative entropy.

We adapt measurement strategies developed for state tomography to universal hypothesis testing by introducing a parameterized tilt into the pretty-good measurement approach. This gives an explicit tilted typical projection under Schur--Weyl decomposition and universally achievable reverse sandwiched Rényi exponent bounds.

Schur--Weyl duality states that the $n$-fold tensor product of $\mathbb{C}^d$ decomposes as 
\begin{align}\label{eq:schur_weyl}
    (\mathbb{C}^d)^{\otimes n} \cong \bigoplus_{\lambda}\Qlambda\otimes\Pc_\lambda,
\end{align}
where $\Qlambda$ is the irreducible representation (irrep) subspace associated with the action of the unitary group $U(d)$ and $\Pc_\lambda$ is the irrep subspace associated with the action of the symmetric group $S_n$. Here, $\lambda$ represents a partition of $n$ into at most $d$ parts, denoted by an array $(\lambda_1, \lambda_2,\ldots,\lambda_d)$ where $ \lambda_1 \geq\lambda_2\geq\cdots\geq \lambda_d\geq 0$ with $  \sum_i\lambda_i=n$, which labels the irreps of both subspaces.
The main idea is that the actions of the unitary group and the symmetric group commute with each other, so they can be simultaneously block-diagonalized. Each block is labeled by the partition $\lambda$.
As shown in \cite[Eqs.~(6.14) and (6.16)]{Harrow}, we can bound their dimensions by
\begin{align}
    &\dim\Qlambda \leq(n+d)^{d(d-1)/2} \label{dimQbound},\\
    (n+d)^{-d(d+1)/2}e^{nH(\bar{\lambda})}\leq&\dim\Pc_\lambda \leq e^{nH(\bar{\lambda})} \label{dimPbound},
\end{align}
where
\begin{align}\label{eq:lambda_bar}
    \bar{\lambda} \coloneqq \operatorname{diag}(\lambda_1/n,\ldots,\lambda_d/n)
\end{align}
is the normalized diagonal matrix associated with the partition $\lambda$, and $H(\bar{\lambda})$ is the Shannon entropy of the normalized partition $\lambda/n$.
The Schur transform is the unitary transformation that implements the decomposition \eqref{eq:schur_weyl}. Applying it to an $n$-fold identical product state yields
\begin{align}
     U_{\mathrm{Sch}}\rho^{\otimes n} U_{\mathrm{Sch}}^{\dagger}=\sum_{\lambda}\ket{\lambda}\bra{\lambda}\otimes q_\lambda^d(\rho)\otimes \mathbb{I}_{\Pc_\lambda},
\end{align}
where $q_\lambda^d(\rho)$ denotes the representation of $\rho$ on $\Qlambda$, and $\{\ket{\lambda}\}$ is an orthonormal basis of label states, such that $\ket{\lambda}\bra{\lambda}$ is the rank-one orthogonal projector of unit trace onto the state $\ket{\lambda}$ of the label register. Because the $n$-fold state is permutation-invariant, it acts as the identity in the permutation subspace. We also define $\Pi_\lambda$ as the projector onto the subspace (block from the block-diagonal matrix) associated with the label $\lambda$:
\begin{align}
\Pi_\lambda\coloneqq U_{\mathrm{Sch}}^\dagger\left[\ket{\lambda}\bra{\lambda}\otimes\mathbb{I}_{\Qlambda}\otimes\mathbb{I}_{\Pc_\lambda}\right]U_{\mathrm{Sch}}.
\end{align}
Direct consequences of this definition are
\begin{align}
    &\Pi_\lambda\sigma^{\otimes n} = \sigma^{\otimes n}\Pi_\lambda = \Pi_\lambda\sigma^{\otimes n}\Pi_\lambda \label{projection commute with state},\\
    &\Pi_\lambda\sigma^{\otimes n}\Pi_\lambda = U_{\mathrm{Sch}}^{\dagger}\left[\ket{\lambda}\bra{\lambda}\otimes q_\lambda^d(\sigma)\otimes \mathbb{I}_{\Pc_\lambda}\right]U_{\mathrm{Sch}}\label{projection lambda property}.
\end{align}
Measuring with the projectors $\{\Pi_\lambda\}$ estimates the spectrum of an unknown state via the normalized partition $\bar{\lambda}$ and is known to be optimal for spectrum estimation~\cite{EstimatingSpectrum, QuantumStateEstimationAndLargeDeviations}. A subsequent measurement on the irrep subspace $\Qlambda$ estimates the eigenbasis by a unitary $U$, yielding the state estimate $\hat{\rho}=U\bar{\lambda}U^\dagger$. To construct our universal test from these estimates, we define, for every pair $(\lambda,U)$ and any $s\in(0,1)$, the tilted measurement operator
\begin{align}\label{eq:tilted_operator}
    M_s(\lambda,U)\coloneqq \frac{\dim \Qlambda}{\operatorname{Tr}\left[q_\lambda^d\left(\bar{\lambda}^{\frac{s}{1-s}}\right)\right]}\Pi_\lambda\left(U\bar{\lambda}^{\frac{s}{1-s}}U^{\dagger}\right)^{\otimes n}\Pi_\lambda.
\end{align}
The tilt reweights the spectrum by the power $s/(1-s)$; this parameter will later generate the reverse sandwiched Rényi relative entropy in the probability bound.

For any pair $(\lambda,U)$, the Born rule gives
\begin{align}
\operatorname{Tr}\left[M_s(\lambda,U)\sigma^{\otimes n}\right]&= 
    \frac{\dim \Qlambda}{\operatorname{Tr}\left[q_\lambda^d\left(\bar{\lambda}^{\frac{s}{1-s}}\right)\right]}\operatorname{Tr}\left[\Pi_\lambda\left(\hat{\rho}^{\frac{s}{1-s}}\right)^{\otimes n}\Pi_\lambda\sigma^{\otimes n}\right]\notag
    \\&=\frac{\dim \Qlambda}{\operatorname{Tr}\left[q_\lambda^d\left(\bar{\lambda}^{\frac{s}{1-s}}\right)\right]}\operatorname{Tr}\left[\Pi_\lambda\left(\hat{\rho}^{\frac{s}{1-s}}\sigma\right)^{\otimes n}\Pi_\lambda\right]\notag
    \\&=\frac{\dim \Qlambda}{\operatorname{Tr}\left[q_\lambda^d\left(\bar{\lambda}^{\frac{s}{1-s}}\right)\right]}\operatorname{Tr}\left[\ket{\lambda}\bra{\lambda}\otimes q_\lambda^d\left(\hat{\rho}^{\frac{s}{1-s}}\sigma\right)\otimes \mathbb{I}_{\Pc_\lambda}\right]\notag
    \\&=\frac{\dim \Qlambda\cdot\dim\Pc_\lambda}{\operatorname{Tr}\left[q_\lambda^d\left(\bar{\lambda}^{\frac{s}{1-s}}\right)\right]}\operatorname{Tr}\left[q_\lambda^d\left(\hat{\rho}^{\frac{s}{1-s}}\sigma\right)\right]. \label{bound_probability}
\end{align}
The second equality uses cyclicity of the trace, the commutation of $\Pi_\lambda$ with $\sigma^{\otimes n}$, and the tensor-product representation; the third follows from \eqref{projection lambda property}; and the last traces out $\ket{\lambda}\bra{\lambda}$ and the permutation subspace, using $\operatorname{Tr}[\ket{\lambda}\bra{\lambda}] = 1$ and $\operatorname{Tr}[\mathbb{I}_{\Pc_\lambda}] = \dim\Pc_\lambda$.
To analyze the resulting error exponent, we must understand quantities of the form
\begin{align} \label{quantity_to_understand}
\operatorname{Tr}\left[q_\lambda^d\left(\rho^{\frac{s}{1-s}}\sigma\right)\right],
\end{align}
where $\rho$ and $\sigma$ are generic quantum states.
The following lemma serves as our main tool.
\begin{lemma}\label{traceboundlemma}
Let $X\succeq 0$ and $k>0$. Then,
\begin{align}
    e^{-nkH(\bar{\lambda})}e^{-nkD(\bar{\lambda} \,\|\, \bar{x})}e^{nk\log\operatorname{Tr}[X]}\leq \operatorname{Tr}\left[q_\lambda^d\left(X^k\right)\right]\leq\dim \Qlambda e^{-nkH(\bar{\lambda})}e^{-nkD(\bar{\lambda} \,\|\, \bar{x}
    )}e^{nk\log\operatorname{Tr}[X]},
\end{align}
where $x$ is the vector of eigenvalues of $X$ and $\bar x = \frac{x}{\operatorname{Tr}[X]}$.
\end{lemma}
\begin{proof}
The proof is included within the proof of \cite[Lemma~2, Section~IV]{SampleOptimalTomographyOfQuantumStates}. The quantity $\operatorname{Tr}[q_\lambda^d(X)]$ has an exact expression in terms of the eigenvalues of $X$, denoted by $x = (x_1,\ldots,x_d)$, via the formula
\begin{align}
    \operatorname{Tr}\left[q_\lambda^d(X)\right] = \sum_{\nu}K_{\lambda\nu}x^\nu,
\end{align}
where $x^\nu \coloneqq x_1^{\nu_1}\cdots x_d^{\nu_d}$, and $K_{\lambda\nu}$ are the Kostka coefficients. Here, the weight vectors $\nu = (\nu_1,\ldots,\nu_d)$ are majorized by the partition $\lambda$ (denoted $\nu \prec \lambda$), meaning that
\begin{align}
\sum_{i=1}^m\nu_i^{\downarrow}\leq \sum_{i=1}^m\lambda_i \quad(1\leq m\leq d),
\end{align}
where $\nu^{\downarrow}$ denotes the components of $\nu$ sorted in descending order. 

To upper-bound this expression, we use $\sum_\nu K_{\lambda\nu} = \dim \Qlambda$. Since majorization implies that $\max_{\nu\prec\lambda}x^{\nu}=x^{\lambda}$ for sorted eigenvalue vectors, the sum is bounded by $\dim\Qlambda$ times its largest term, attained by pairing the largest eigenvalue $x_1$ with the largest possible exponent $\nu_1=\lambda_1$, then $x_2$ with $\nu_2=\lambda_2$, and so on:
\begin{align}
\operatorname{Tr}\left[q_\lambda^d(X)\right] \leq\dim \Qlambda\cdot x_1^{\lambda_1} \cdots x_d^{\lambda_d}.
\end{align}

For powers of positive matrices  $X^k$, the corresponding eigenvalues are given by $x^k = (x_1^k,\ldots,x_d^k)$. Applying the upper bound yields
\begin{align}
\operatorname{Tr}\left[q_\lambda^d(X^k)\right]  \leq\dim \Qlambda \cdot x_1^{k\lambda_1}\cdots x_d^{k\lambda_d}.\label{eq:trace_qk_upper}
\end{align}
We can rewrite this expression as exponential terms,
\begin{align}
x_1^{k\lambda_1}\cdots x_d^{k\lambda_d} &= e^{k\sum_i\lambda_i\log x_i} \notag\\
    &= e^{nk\sum_i\bar{\lambda}_i \log \bar{\lambda}_i -nk\sum_i\bar{\lambda}_i \log \bar{\lambda}_i+nk\sum_i\bar{\lambda}_i\log x_i} \notag\\
    &=e^{-nkH(\bar{\lambda})}e^{-nk\Big(\sum_i \bar{\lambda}_i\log \bar{\lambda}_i-\sum_i\bar{\lambda}_i\log \left(\frac{x_i}{\operatorname{Tr}[X]}\right)-\sum_i\bar{\lambda}_i\log\operatorname{Tr}[X]\Big)} \notag\\
    &=e^{-nkH(\bar{\lambda})}e^{-nkD(\bar{\lambda} \,\|\, \bar x)} e^{nk\log\operatorname{Tr}[X]}.\label{eq:exp_rewrite}
\end{align}

To lower-bound $\operatorname{Tr}\left[q_\lambda^d(X^k)\right]$, we use that it is a sum of nonnegative terms, so we can lower-bound it by its largest term:
\begin{align}\label{eq:trace_qk_lower}
\operatorname{Tr}\left[q_\lambda^d\left(X^k\right)\right] \geq e^{-nkH(\bar{\lambda})}e^{-nkD(\bar{\lambda} \,\|\, \bar x)} e^{nk\log\operatorname{Tr}[X]}.
\end{align}
The lemma thus follows from \eqref{eq:trace_qk_upper}, \eqref{eq:exp_rewrite}, and \eqref{eq:trace_qk_lower}.
\end{proof}
Applying Lemma~\ref{traceboundlemma} to the trace in \eqref{bound_probability} gives the main estimate used in the derivation of the achievable exponent:
\begin{lemma}\label{lemma:outcome_bound}
Consider the measurement operator given in \eqref{eq:tilted_operator}. Then,
\begin{align}
\operatorname{Tr}[M_s(\lambda,U)\sigma^{\otimes n}] \leq (n+d)^{d(d-1)}e^{-nD_{R,s}\big(U\bar{\lambda}U^{\dagger}\,\|\,\sigma\big)},
\end{align}
where $D_{R,s}$ is defined in \eqref{reverse_relative_entropy}.
\end{lemma}
\begin{proof}
Let $\hat{\rho}=U\bar\lambda U^\dagger$. Apply Lemma~\ref{traceboundlemma} with
\[
    X=\left(\hat{\rho}^{\frac{s}{2(1-s)}}\sigma\hat{\rho}^{\frac{s}{2(1-s)}}\right)^{1-s},
    \qquad k=\frac{1}{1-s},
\]
so that $X^k=\hat{\rho}^{\frac{s}{2(1-s)}}\sigma\hat{\rho}^{\frac{s}{2(1-s)}}$. By cyclicity of the trace and the representation property of $q_\lambda^d$, the right-most trace on the right-hand side of \eqref{bound_probability} can be written as $\operatorname{Tr}[q_\lambda^d(X^k)]$. The upper bound of Lemma~\ref{traceboundlemma} further gives
    \begin{align}
        \operatorname{Tr}\left[q_\lambda^d\left(X^k\right)\right]
        &\leq \dim \Qlambda e^{-n\frac{1}{1-s}H(\bar{\lambda})}e^{-n\frac{1}{1-s}D(\bar{\lambda} \,\|\, \bar{x})}
        e^{n\frac{1}{1-s}\log\operatorname{Tr}[X]}.\label{bound3}
    \end{align}
     Furthermore, by using the lower bound of Lemma~\ref{traceboundlemma} for $X=\bar\lambda$ and $k=\frac{s}{1-s}$, we can bound the denominator on the right-hand side of \eqref{bound_probability} by 
    \begin{align}\label{bound_denominator}
        \frac{1}{\operatorname{Tr}[q_\lambda^d(\bar\lambda^{\frac{s}{1-s}})]} \leq e^{n\frac{s}{1-s}H(\bar\lambda)} e^{n\frac{s}{1-s} D(\bar\lambda\,\|\,\bar\lambda)} e^{-n\frac{s}{1-s}\log\operatorname{Tr}[\bar\lambda]} = e^{n\frac{s}{1-s}H(\bar\lambda)}.
    \end{align}
By combining \eqref{bound3} and \eqref{bound_denominator} with \eqref{bound_probability}, we thus get
\begin{align}
   \operatorname{Tr}\left[M_s(\lambda,U)\sigma^{\otimes n}\right]
   &= \frac{\dim \Qlambda\cdot\dim\Pc_\lambda}{\operatorname{Tr}\left[q_\lambda^d\left(\bar{\lambda}^{\frac{s}{1-s}}\right)\right]}\operatorname{Tr}\left[q_\lambda^d\left(\hat{\rho}^{\frac{s}{1-s}}\sigma\right)\right]\notag\\
   &\leq (\dim \Qlambda )^2\cdot\dim\Pc_\lambda e^{n\frac{s}{1-s}H(\bar\lambda)} e^{-n\frac{1}{1-s}H(\bar{\lambda})}e^{-n\frac{1}{1-s}D(\bar{\lambda} \,\|\, \bar{x})}e^{n\frac{1}{1-s}\log\operatorname{Tr}[X]}\notag\\
   &\leq (\dim \Qlambda )^2\cdot e^{nH(\bar\lambda)} e^{-nH(\bar\lambda)} e^{n\frac{1}{1-s}\log\operatorname{Tr}[X]}\notag
    \\ &\leq(n+d)^{d(d-1)} e^{-nD_{R,s}\left(\hat{\rho}\,\|\,\sigma\right)},
    \end{align}
    where we also used \eqref{dimQbound}, \eqref{dimPbound}, and the fact that the factor involving $D(\bar\lambda\,\|\,\bar x)$ is at most one. The remaining exponential is $e^{-nD_{R,s}(\hat{\rho}\,\|\,\sigma)}$ by the definition \eqref{reverse_relative_entropy} of the reverse sandwiched Rényi relative entropy.
\end{proof}
We next define the typical projection
\begin{align}
    \Pi^n_{\rho,\delta,s} \coloneqq \sum \int_{(\lambda,U)\in\Bc_{\delta,s}(\rho)} M_{s}(\lambda, U)   \, \mathrm{d}U,
\end{align}
where $\Bc_{\delta,s}(\rho) = \{(\lambda,U): D_{R,s}(U\bar{\lambda}U^{\dagger}\,\|\,\rho)\leq\delta\}$. This set contains outcomes whose estimates are close to the null state in reverse sandwiched Rényi relative entropy. The operator $\Pi^n_{\rho,\delta,s}$ defines a POVM, since it is an integral of positive operators, and it accepts the null state with high probability, as demonstrated in the following lemma.
\begin{lemma}\label{lem:typicality}
    For every $\delta>0$ and every positive integer $n$, we have
\begin{align}
    \operatorname{Tr}[\Pi^n_{\rho,\delta,s}\rho^{\otimes n}] \geq 1-(n+d)^{d^2}e^{-n\delta}.
\end{align}
\end{lemma}
\begin{proof}
The probability of rejecting the null state is obtained by integrating over outcomes in the complement of $\Bc_{\delta,s}(\rho)$. We then have
    \begin{align}
        \operatorname{Tr}[(\mathbb{I}-\Pi^n_{\rho,\delta,s})\rho^{\otimes n}] &= \sum\int_{(\lambda,U)\in \Bc_{\delta,s}^c(\rho)} \operatorname{Tr}[M_{s}(\lambda,U)\rho^{\otimes n}] \, \mathrm{d}U \notag
        \\&\leq (n+d)^{d(d-1)}\sum\int_{(\lambda,U)\in \Bc_{\delta,s}^c(\rho)} e^{-nD_{R,s}(U\bar{\lambda}U^{\dagger} \,\|\, \rho)} \, \mathrm{d}U \notag
        \\&\leq (n+d)^{d(d-1)}e^{-n\delta}\sum\int_{(\lambda,U)\in \Bc_{\delta,s}^c(\rho)}  \, \mathrm{d}U \notag
        \\&\leq (n+d)^{d(d-1)}e^{-n\delta}\sum\int \, \mathrm{d}U \notag
        \\&\leq (n+d)^{d(d-1)}e^{-n\delta}(n+1)^d \leq (n+d)^{d^2}e^{-n\delta},
    \end{align}
     where the first inequality follows from Lemma~\ref{lemma:outcome_bound}; the second inequality follows because $D_{R,s}(U\bar{\lambda}U^{\dagger}\,\|\,\rho)\geq\delta$ outside the typical set $\Bc_{\delta,s}(\rho)$; the third inequality follows by extending the integration domain to all possible outcomes; and the last line follows because there are at most $(n+1)^d$ partitions $\lambda$ and $\int\mathrm{d}U=1$, so the total measure is at most $(n+1)^d\leq(n+d)^d$.
\end{proof}

We are now ready to prove Theorem~\ref{main_result_achievability}. For fixed $s\in(0,1)$, take
\[
    T_n=\Pi^n_{\rho,\delta_n,s},\qquad
    \delta_n=\frac{d^2+1}{n}\log(n+d).
\]
The two outcomes are $T_n$ (accept the null) and $\mathbb{I}-T_n$ (reject it). Lemma~\ref{lem:typicality} gives the type-I upper bound
\begin{align}
    \alpha_n(\rho\mid T_n)=\operatorname{Tr}[(\mathbb{I}-T_n)\rho^{\otimes n}]
    \leq (n+d)^{d^2}e^{-n\delta_n}=\frac{1}{n+d}\longrightarrow0.
\end{align}
For the type-II error, we apply Lemma~\ref{lemma:outcome_bound} directly to each accepted outcome to obtain
\begin{align}
    \beta_n(\sigma\mid T_n)
    &=\sum\int_{(\lambda,U)\in\Bc_{\delta_n,s}(\rho)}
       \operatorname{Tr}[M_s(\lambda,U)\sigma^{\otimes n}]\,\mathrm{d}U\notag\\
    &\leq (n+d)^{d(d-1)}
       e^{-n\inf_{(\lambda,U)\in\Bc_{\delta_n,s}(\rho)}
       D_{R,s}(U\bar{\lambda}U^\dagger\,\|\,\sigma)}
       \sum\int_{\Bc_{\delta_n,s}(\rho)}\mathrm{d}U\notag\\
    &\leq (n+d)^{d^2}
       e^{-n\inf_{(\lambda,U)\in\Bc_{\delta_n,s}(\rho)}
       D_{R,s}(U\bar{\lambda}U^\dagger\,\|\,\sigma)}.\label{eq:beta_n_upper_bound}
\end{align}
In the last step, we use again the fact that there are at most $(n+1)^d$ partitions $\lambda$ and that $\int\mathrm{d}U=1$. Taking logarithms on both sides of \eqref{eq:beta_n_upper_bound} and dividing by $-n$ then gives
\begin{align}\label{eq:exponent_lower_bound}
    -\frac1n\log\beta_n(\sigma\mid T_n)
    \geq -\frac{d^2}{n}\log(n+d)
       +\inf_{(\lambda,U)\in\Bc_{\delta_n,s}(\rho)}
       D_{R,s}(U\bar{\lambda}U^\dagger\,\|\,\sigma).
\end{align}
As $\delta_n\to0$, the accepted estimates concentrate around $\rho$. Since, for a fixed $s\in(0,1)$, the reverse sandwiched Rényi relative entropy is jointly continuous in its two arguments, we obtain that
\begin{align}
    \lim_{n\to\infty}
    \inf_{(\lambda,U)\in\Bc_{\delta_n,s}(\rho)}
       D_{R,s}(U\bar{\lambda}U^\dagger\,\|\,\sigma)
    =D_{R,s}(\rho\,\|\,\sigma).
\end{align}
The polynomial term on the right-hand side of \eqref{eq:exponent_lower_bound} vanishes because $\frac{d^2}{n}\log(n+d)\to0$. Thus, for every $s\in(0,1)$, the construction guarantees an exponent of at least $D_{R,s}(\rho\,\|\,\sigma)$. Since $s\in(0,1)$ was arbitrary, taking the limit $s\to1^-$ further gives
\begin{align}
    \lim_{s\to1^-}D_{R,s}(\rho\,\|\,\sigma)=D_R(\rho\,\|\,\sigma).
\end{align}

\appendix
\section{Derivation of the Coefficient $\gamma(x,y,z)$} \label{derivation_appendix}
We provide here an explicit derivation of the coefficient $\gamma(x,y,z)$ in \eqref{coefficient_lambda_xyz}. A generic full-rank qubit state with Bloch vector $\vec{r}=(r_x,r_y,r_z)$ can be written as
\begin{align}
    \omega = \frac{1}{2}(\mathbb{I}+r_x X + r_y Y + r_z Z), \quad  0 < |\vec{r}| = t < 1,
\end{align}
where $X,Y,Z$ denote the standard Pauli matrices. Functional calculus gives
\begin{align}
    \log\omega = c(t)\mathbb{I}+\frac{\arctanh{(t)}}{t}(r_x X + r_y Y + r_z Z) \label{functional_calculus_identity},
\end{align}
where
\begin{align}
    c(t) = \frac{1}{2}\log \frac{1-t^2}{4}.
\end{align}
The states in the two-dimensional construction have Bloch vectors
\begin{align}
   \vec r_1=(x,y,z),\quad  \vec r_2=(-x,-y,z),\quad t=\sqrt{x^2+y^2+z^2}.
\end{align}
Applying \eqref{functional_calculus_identity} to $S_1$ and $S_2$, the $X$ and $Y$ terms cancel out in the average, and hence
\begin{align}
    \frac{\log S_1 +\log S_2}{2}=c(t)\mathbb{I}+\frac{z}{t}\arctanh{(t)}Z \label{average_term}.
\end{align}
It remains to compute the coefficient of $Z$ in $\log (S_1 \# S_2)$. Let 
\begin{align}
    G\coloneqq S_1 \# S_2.
\end{align}
Since $\Lambda(S_1,S_2)$ is diagonal and \eqref{average_term} is diagonal, $\log G$ is diagonal, and hence $G$ is diagonal. Write
\begin{align}
    G=\begin{pmatrix}
        g_0 & 0 \\ 0 & g_1
    \end{pmatrix}.
\end{align}
The geometric mean $G= A \# B$ is the unique positive definite solution of 
\begin{align}
    GB^{-1}G = A.
\end{align}
Thus $G = S_1 \# S_2$ satisfies
\begin{align}
    GS_2^{-1}G= S_1 \label{equation_geometric_mean}.
\end{align}
Since
\begin{align}
    S_2^{-1}=\frac{2}{1-t^2}\begin{pmatrix}
        1-z&x-iy\\x+iy&1+z
    \end{pmatrix},
\end{align}
\eqref{equation_geometric_mean} becomes 
\begin{align}
    \frac{2}{1-t^2} \begin{pmatrix}
        g_0^2(1-z) &g_0g_1(x-iy)\\g_0g_1(x+iy)&g_1^2(1+z)
    \end{pmatrix}
    = \frac{1}{2} \begin{pmatrix}
        1+z & x-iy\\x+iy & 1-z
    \end{pmatrix}.
\end{align}
We do not need the full diagonal solution for $G$. Dividing the two diagonal equations gives
\begin{align}
    \left(\frac{g_0}{g_1}\right)^2=\left(\frac{1+z}{1-z}\right)^2.
\end{align}
Since $g_0,g_1>0$,
\begin{align}
    \frac{g_0}{g_1}=\frac{1+z}{1-z}.
\end{align}
We can divide $\log G$ into a scalar part and a traceless part. The scalar part cancels out when forming $\Lambda(S_1,S_2)$, and the traceless part is
\begin{align}
    \log G - \operatorname{Tr}[\log G]\frac{\mathbb{I}}{2} = \frac{1}{2}(\log g_0 - \log g_1)Z = \arctanh(z)\,Z. \label{traceless_part}
\end{align}
Finally, subtracting the $Z$ coefficient in \eqref{average_term} from \eqref{traceless_part} gives
\begin{align}
    \Lambda(S_1,S_2) = \left(\arctanh(z)-\frac{z}{t}\arctanh(t)\right)Z.
\end{align}
It follows that
\begin{equation}
    \gamma(x,y,z) = \arctanh(z)-\frac{z}{t}\arctanh(t).
\end{equation}

\section*{Acknowledgments}
Stimulating discussions with Enrique Cervero during his visit to UC3M are gratefully acknowledged.

Generative AI tools, namely OpenAI Codex (GPT-5.5 and GPT-6 models), Google Antigravity (Gemini 4 models), and GitHub Copilot, were used to assist the authors with language editing, \LaTeX{} formatting, and searching for related references (which were independently read and verified by the authors). These tools also assisted in the proof of Theorem~\ref{thm:tomography_to_quht}, in the derivation of the coefficient $\gamma(x,y,z)$ in Appendix~\ref{derivation_appendix}, and in writing the code for the numerical example in Section~\ref{sec:numerical_example}. All mathematical arguments, proofs, code, and written text were carefully verified and revised by the authors.

\bibliographystyle{IEEEtran}
\bibliography{biblio}

\end{document}